\documentclass[showpacs,twocolumn,preprintnumbers,prl,amssymb]{revtex4}
\usepackage{graphicx}
\usepackage{dcolumn}
\usepackage{bm}
\usepackage{amsmath}
\usepackage{amsfonts}
\usepackage{amssymb}
\usepackage{ulem}
\usepackage{color}

\begin{document}

\title{Anomalous Doping Effects on Charge Transport in Graphene Nanoribbons}

\author{Blanca Biel$^{1,3}$, X. Blase$^2$, Fran\c cois Triozon$^1$ and Stephan Roche$^3$}

\affiliation{ 
$^1$ CEA, LETI, MINATEC, F38054 Grenoble, France.\\
$^2$ Institut N\'{e}el, CNRS and Universit\'{e} Joseph Fourier,
B.P. 166, 38042 Grenoble Cedex 09, France.\\
$^3$ CEA, Institut of Nanosciences and Cryogenics, INAC/SPSMS/GT  17 rue des Martyrs, 38054 Grenoble Cedex 9, France
}

\date{\today}

\begin{abstract}
We present first-principles calculations of quantum transport in chemically
doped graphene nanoribbons
with a width of up to 4 nm.
The presence of boron and nitrogen impurities is shown to yield resonant backscattering, whose
features are strongly dependent on the symmetry and the width of the ribbon,
as well as the position of the dopants.
Full suppression of backscattering is obtained on the $\pi-\pi^*$ \textsl{plateau}
when the impurity preserves the mirror
symmetry of armchair ribbons. Further, an unusual acceptor-donor transition
is observed in zig-zag ribbons.
These unconventional doping effects could
be used to design novel types of switching devices.
\end{abstract}

\maketitle
The ability to single out a single graphene plane, through an exfoliation process \cite{graphene_exf},
or by means of epitaxial growth \cite{graphene_epit}, has opened novel 
opportunities for exploring low dimensional transport in a material with remarkable
 electronic properties \cite{RMP}. Additionally,
the development of graphene-based nanoelectronics has attracted much attention owing to the
promising large scale integration expectations \cite{graphene_epit,graph_transistor}. However,
with 2D graphene being a zero-gap semiconductor, its use in an
active electronic device such as a field effect transistor (FET) requires
a reduction of its lateral size to benefit from quantum confinement effects.
Graphene nanoribbons (GNRs) are strips of graphene whose 
electronic properties depend on their edge symmetry and width \cite{Ribbonpure},
and can be either patterned by plasma etching \cite{IBM,KimENRJGAP}
or derived chemically \cite{Dai}. Band-gap engineering of GNRs
has been experimentally demonstrated \cite{KimENRJGAP}, 
and GNRs-based FET with a width of several tens
of nanometers down to 2 nm have been characterized \cite{Dai}. 

Chemical doping aims at producing {\bf\it p}-doped or {\bf\it n}-doped transistors,
which are crucial for building logic functions and complex circuits \cite{4}. Doping also allows new applications
 such as chemical sensors, or electrochemical switches \cite{MOL}. In carbon-based materials
{\bf\it p}-type ({\bf\it n}-type) doping can be achieved by boron (nitrogen) atom
substitution within the carbon matrix \cite{BdopedCVD}.
For metallic carbon nanotubes (CNTs), Choi
and co-workers reported that boron (B) and nitrogen (N) impurities
yield quasibound states which strongly
backscatter propagating charge for specific resonance energies
\cite{5}. The interplay between those resonance
energies and external parameters (electric or magnetic fields) may also enable the design of novel
kinds of CNT-based switching devices \cite{8}, whereas the 
existence of quasibound states
related to topological defects was unveiled by STM measurements \cite{17}.

In this Letter, we report on an \textsl{ab initio}
study of the effect of both {\bf\it p}-type (B) and {\bf\it n}-type (N) doping 
on quantum transport in GNRs with widths within
the experimental scope. In contrast with CNTs,
doping in GNRs turns out to display more complex features
 depending on the dopant position, ribbon width and symmetry.
Theoretically, two types of GNRs with
highly symmetric edges have been described,
namely zig-zag (zGNRs) and armchair (aGNRs) \cite{Ribbonpure}.
Some recent works have reported
on the effect of doping in extremely narrow zGNRs \cite{dopedGNRs},
mostly with a width in order of 1 nm.

Following Ref. \cite{Louie-gaps}, we refer to a zGNR (aGNR) with N zig-zag chains (dimers) 
contained in its unit cell as 
a N-zGNR (N-aGNR). We have studied aGNRs
and zGNRs with widths
between 2.3 and 4.2 nm, already within the current experimental scope \cite{Dai, Lambin}. 
The scattering potential around the dopant is obtained using first principles calculations
(SIESTA code \cite{SIESTA}) within the local density approximation \cite{Troullier,NOTE0}.

\begin{figure}[htbp]
 \centering
 \includegraphics[width=7.5cm]{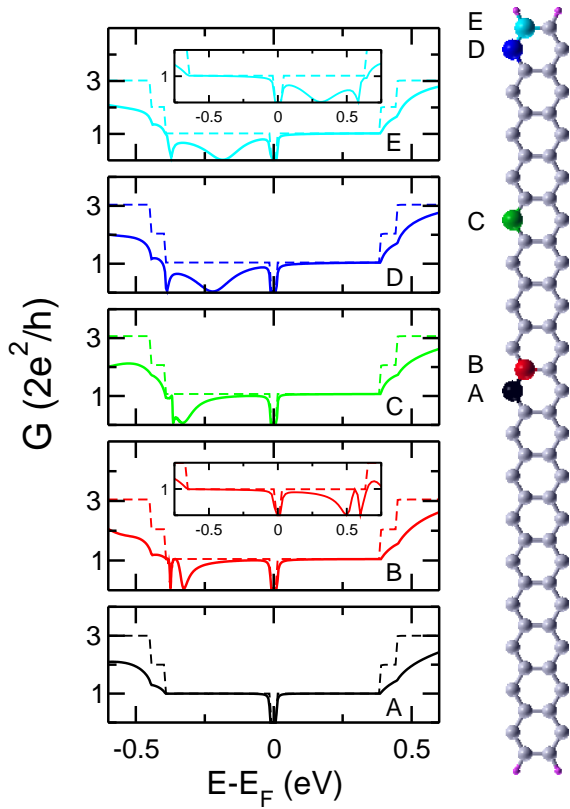}
 \caption{(color online). Left: Conductance of the 35-aGNR as a function of energy for different B
positions. The dashed lines correspond to conductance for the undoped case.
Insets: same as in main frame for two selected nitrogen dopant positions
(at the edge (E), and off-center (B)) for the 20-aGNR.
Right: unit cell of the 35-aGNR showing the considered dopant positions.
The carbon atoms are shown in grey and the passivating H atoms in pink;
the other colored atoms represent B or N atoms, each color referring to a corresponding colored solid line 
conductance curve.}
 \label{FIG1}
\end{figure}
We start with the case of aGNRs. \textsl{Ab initio} studies show that aGNRs are always semiconducting 
\cite{Louie-gaps,Scuseria-gaps,White-gaps} with width-dependent bandgap scaling.
We have studied the 20-, 34- and 35-aGNRs, with
widths of 2.3, 4.0 and 4.2 nm, respectively.
In carbon nanotubes, two acceptor (donor) quasibound states have been predicted
below (above) the charge neutrality point (CNP) at low energy values when a single carbon atom
 is substituted by a boron (nitrogen) impurity \cite{5,Mahan}. 
 
However, in contrast with CNTs, the energies of the quasibound states in GNRs
are strongly dependent on the position of the impurity with respect to the GNR edges. A clear increase in binding energy of the
bound state associated with the broad drop in the conductance in Fig. \ref{FIG1}
is observed as the dopant approaches the edge. The large variation of resonant energies with dopant position indicates 
that random distribution of
impurities will lead to a rather uniform reduction of conductance over the occupied-states
part of the first conduction \textsl{plateau}. This is in sharp contrast to the case of CNTs,
where resonant energies
do not depend on the position of the dopant around the tube circumference.
Our results for the 34-aGNR (not shown)
confirm this behavior also for semiconducting aGNRs.
\begin{figure}[htbp]
\centering
 \includegraphics[width=6.0cm]{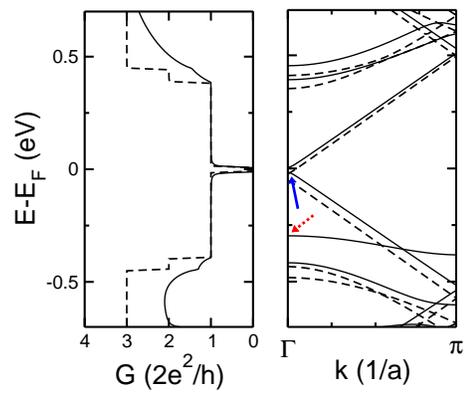}
 \caption{(color online). Conductance as a function of the energy (left) and bandstructure (right)
of the 35-aGNR \cite{NOTE1}. Dashed lines correspond to the undoped ribbon;
solid lines correspond to the case of B at the center.
The solid blue (dashed red) arrow shows
the first (second) band below the charge neutrality point (CNP) for the doped case.
}
\label{FIG2}
\end{figure}
\begin{figure*}[htbp]
\centering
 \includegraphics[width=16.0cm]{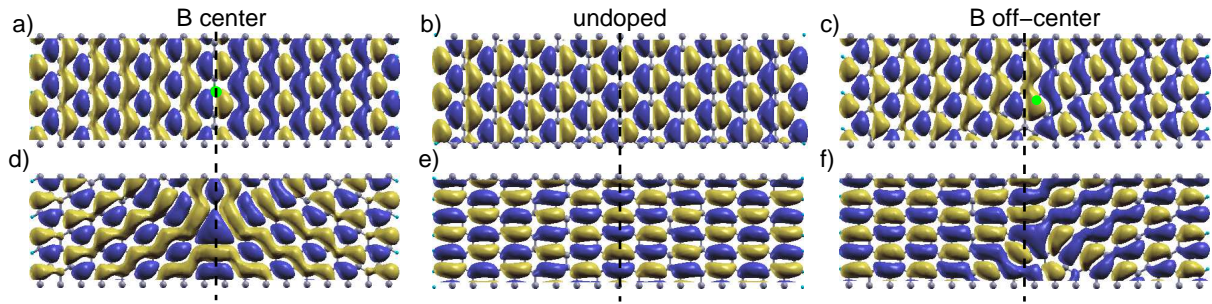}
 \caption{(color online). a-c) 2-dimensional projection of the wavefunctions at the $\Gamma$ point associated
 with the first occupied band below the CNP (solid blue arrow in Fig. \ref{FIG2}),
for a 35-aGNR with B at center (a), no dopant (b), and B-off center(c).
d-f) Same as in a-c) for the second occupied band below the CNP (dashed red arrow in Fig. \ref{FIG2}).
The position of the B impurity is marked in green.
Blue or yellow color corresponds to opposite sign of the wavefunction.
Dashed lines show the ribbon axis (located exactly in the middle of the ribbon). 
}
 \label{FIG3}
\end{figure*}
In addition, for certain dopant positions, symmetry effects yield a full suppression
of backscattering even in the presence of bound states.
Indeed, when B is placed at the exact center of the ribbon,
the transmission at the first \textsl{plateau} is found to be insensitive
to the presence of the impurity (Fig. \ref{FIG1}-left, bottom curve). 
Conversely,
as the defect approaches the ribbon edge,
the energy resonance floats up towards the CNP and
the conductance is clearly degraded by
the quasibound states induced by the impurity (Fig. \ref{FIG1}-left).

To understand such phenomena, we must first note that
GNRs, unlike CNTs, do not always present a well defined
parity associated to mirror reflections with respect to their axis.
An ideal odd-index aGNR retains a 
single mirror symmetry plane
(perpendicular to the plane of the ribbon and containing the ribbon axis), 
and its eigenstates will thus present a well-defined parity
with respect to this symmetry plane (see Figs. \ref{FIG3}b and \ref{FIG3}e).
The eigenstates of the doped ribbon keep the same parity with respect to this mirror plane
provided that the potential induced by the dopant
preserves this symmetry \cite{5,Kim}. For the case of an odd-index aGNR, this can only occur
when the dopant is located exactly at the central dimer line, as is illustrated in
Figs. \ref{FIG3}a and \ref{FIG3}d.
 \begin{figure*}[htbp]
  \centering
  \includegraphics[width=6.0cm]{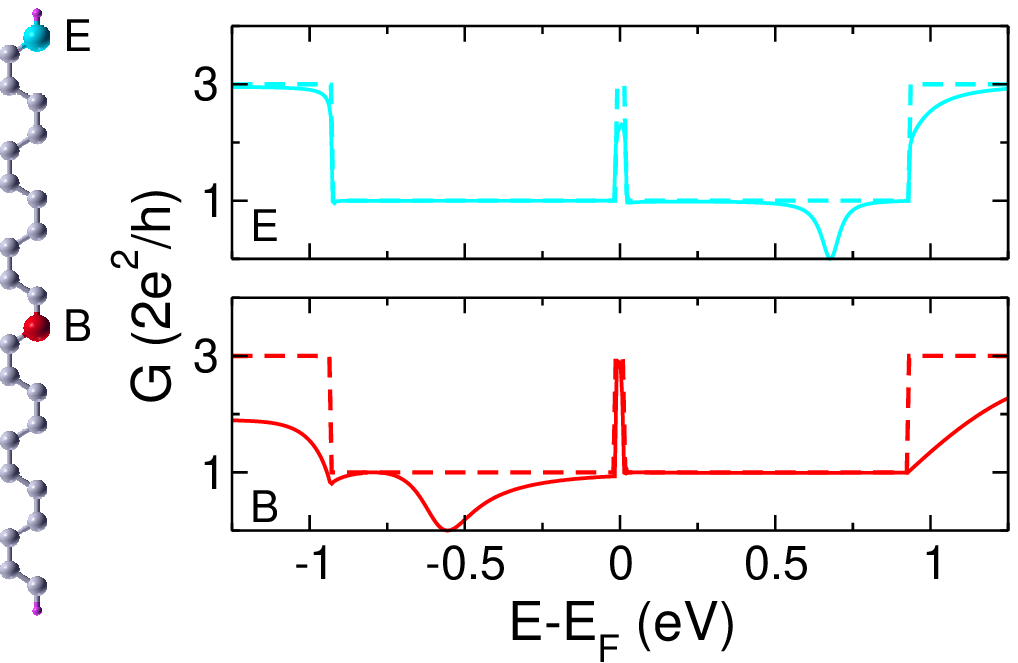}
 \hspace{.8cm}
  \includegraphics[width=6.0cm]{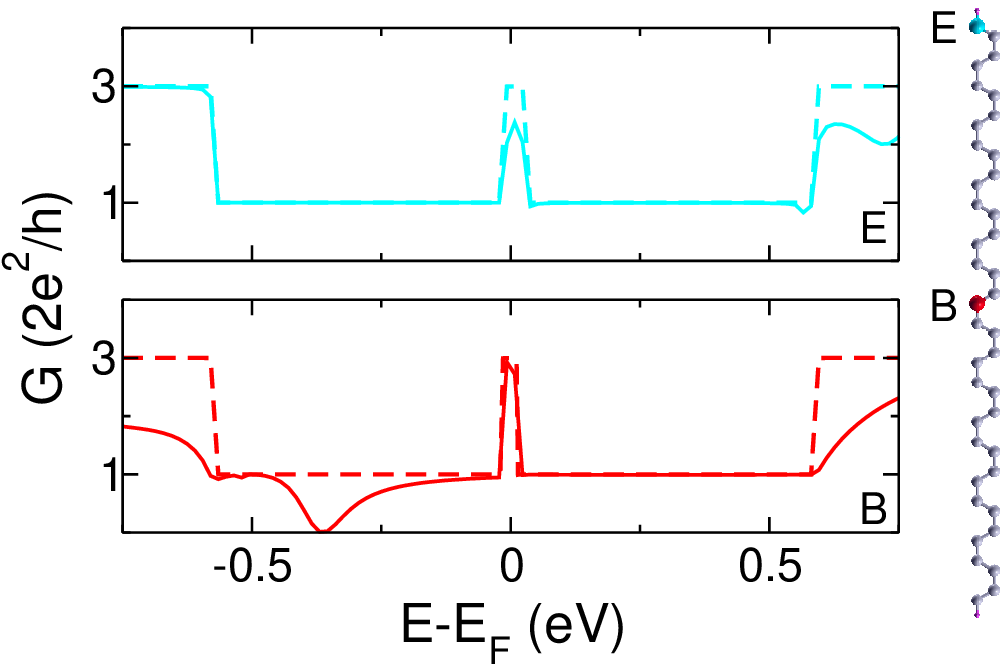}
  \caption{(color online). Conductance as a function of the energy for the 12-zGNR (left) and
20-zGNR (right) with a single B dopant placed
at the ribbon edge (E, top panel) and off-center (B, bottom panel).
The dashed lines correspond to the undoped case. The red or light blue colored atoms
in the unit cell show the position of the dopant for each ribbon.}
  \label{FIG4}
 \end{figure*}
Comparing the wavefunction at the $\Gamma$ point
associated to the first band below the CNP for both the ideal ribbon (Fig. \ref{FIG3}b) and the doped
(B at center) one (Fig. \ref{FIG3}a),
we observe that states around the CNP are only weakly affected by the
impurity, indicating that there is no mixing with neighboring bands (which present an opposite parity).
As a result, backscattering does not occur
despite the presence of impurity states with energy values within the first
\textsl{plateau} (Fig. \ref{FIG2}, dashed red arrow, and Fig. \ref{FIG3}d).
For any other position of the dopant, the well-defined
parity of the wavefunctions will not be preserved, as is shown in Figs. \ref{FIG3}c and \ref{FIG3}f
for the B off-center case (red atom (B) in Fig. \ref{FIG1}). 
In this case, coupling between all states restores 
backscattering efficiency, yielding a full suppression of the single available 
conduction channel
at a certain resonance energy determined by
the precise location of the dopant 
(red curve (B) in Fig. \ref{FIG1}).

In comparison with B-doping, the impact of N impurities on the ribbon transport properties 
manifests in a close symmetric fashion with respect to the CNP. This is illustrated in
the insets of Fig. \ref{FIG1}, where the conductances for a 20-aGNR 
with N off-center (B) and at edge (E) are shown. The same symmetry
considerations aforementioned also apply in the case of odd-index N-doped aGNRs.

We consider now the case of doped zGNRs, and present two different systems:  
namely a 12-zGNR ($\approx$ 2.4 nm width) and a 20-zGNR ($\approx$ 4.1 nm width). 
zGNRs are known to display
very peculiar electronic properties, with wavefunctions
 sharply localized along the GNRs
edges at low energies, which significantly
affect their transport properties \cite{White-transport}. Simple tight-binding 
models found that zGNRs are
always metallic with the presence of sharply localized edge states 
in the vicinity of the Fermi level \cite{Ribbonpure}, whereas spin-dependent 
\textsl{ab initio} calculations report on a small bandgap opening up \cite{Louie}. 
Here spin is neglected,
and we focus on the effect of a boron defect on the transport properties.

Fig. \ref{FIG4} presents the conductance for B-doped 12-zGNR (left) and 20-zGNR (right).
In contrast to aGNRs, where the acceptor character of boron is maintained 
regardless of the position of the impurity, B-doped zGNRs exhibit both acceptor 
and, more unexpectedly, donor character when the dopant is
placed either at the center or at the edges, respectively.
This effect may be related to the competition 
between two different phenomena,
namely the Coulomb interaction of charge carriers with
the ion impurity and correlation between charges at the edges. Our results 
for the narrower 12-zGNR B-doped ribbon confirm such
prior prediction \cite{Ndoped-zz} for N-doped zGNRs
\cite{NOTE2}.
However, for the wider 20-zGNR, a defect located at the edge
has almost no impact on the conduction efficiency in the \textsl{plateau} around the CNP.
This is in striking contrast to the results for the aGNRs, where doping has its maximum effect 
when the impurity is placed at the edge. This suggests that
modification of the electronic properties of zGNRs solely by means of edge doping or functionalization might not
be significant on zGNRs wider than a few nanometers. On the other hand, the geometry of symmetric (even index) zGNRs does not allow a symmetry axis going through a single impurity \cite{Wang}. As a result, the suppression of backscattering observed in the case of aGNRs cannot take place here.
  
In conclusion, \textsl{ab initio} calculations of charge transport in
boron and nitrogen doped GNRs with a width of up to 4.2 nm have been reported. Doping effects depending on the ribbon symmetry and width were unveiled, such as a full suppression of backscattering for symmetry-preserving impurity potentials in armchair ribbons, and upward energy shift for the quasibound state resonances in both armchair and zigzag ribbons. Such predictions could be experimentally confirmed, following recent progress achieved on the STM exploration of graphene edges \cite{Enoki} and on boron-doped oriented pyrolytic graphite \cite{EnokiB}. Finally, chemical doping could be used to enlarge the bandgap of a fixed GNR width, resulting in the enhancement of device performances \cite{Bielnp}. 

This work was financially supported by the GRAPHENE project
of CARNOT Institute (LETI), the European GRAND project (ICT/FET)
and the ANR-06-NANO-069-02 ACCENT.
The authors are indebted to Ch. Adessi and Y.-M. Niquet for technical support
on TABLIER and CEA/TB\_Sim \textsl{ab initio} transport codes.
We thank the CEA/CCRT supercomputing facilities for computational resources. Discussions with A. Cresti are acknowledged.


\begin{thebibliography}{99}
\bibitem{NOTE0}
Periodic boundary conditions are used, with supercells
long enough
(formed by 10 unit cells (u.c.) for aGNRs and by 17 u.c. for zGNRs, that is, up to 40 \AA
\ in length)
to prevent interaction between impurities
in different supercells. Consequently, 
only the $\Gamma$ point is included
in the reciprocal space sampling along the
nanoribbon axis direction.
We have checked that the size of the supercell and
of the (localized) basis (double-$\zeta$) provides well converged conductance spectra.
The ribbon edges have been passivated with H atoms to saturate
the dangling $\sigma$ bonds of the edge
C atoms \cite{Mauri}. Each supercell contains a single dopant, 
that has been placed at different
positions across the direction transversal to the ribbon axis for each relaxation process. 
The initial structures have been relaxed until the forces were smaller than 0.04 eV/\AA.
The transport calculations are based on a first principles
implementation of the Landauer-B\"uttiker approach (for instance see \cite{Adessi}).
\bibitem{NOTE1}
The bandstructure has been calculated for a supercell geometry
containing 10 unit cells and a single dopant (with a total number of 740 atoms in the supercell).
\bibitem{NOTE2}
With the difference that the acceptor character of boron with respect to carbon
will create quasibound states symmetrically positioned with respect to the CNP
to those caused by donor-like N impurities.
\end{thebibliography}
\end{document}